\documentclass[a4paper,11pt]{article}
\pdfoutput=1 

\usepackage{jheppub} 

\usepackage[T1]{fontenc} 

\usepackage[all]{xy}

\def\be{\begin{equation}}
\def\ee{\end{equation}}
\def\ba{\begin{eqnarray}}
\def\ea{\end{eqnarray}}

\def\rd{\mathrm{d}}
\def\rD{{\rm D}}
\def\CL{{\cal L}}
\def\g{\mathfrak{g}}
\def\Lieg{{\mathrm{Lie}(G)}}
\def\CG{{\cal G}}
\def\gr{\mathrm{g}}
\def\h{\mathfrak{h}}
\def\CM{{\cal M}}
\newcommand{\md}{\mathrm{d}}
\def\tH{\widetilde{H}}
\def\tQ{\widetilde{Q}}
\def\te{\widetilde{\epsilon}}
\def\tCM{\widetilde{\CM}}

\let\a=\alpha \let\b=\beta

\let\m=\mu  \let\S=\Sigma
\let\r=\rho

\newcommand*{\R}{{\mathbb R}}
\def\tr{{\it tr\/}}

\newcommand {\subplus}{\mathop{{\subset}\llap{\raise
0.15pt\hbox{\normalfont\small+}\hskip 0.5pt}}}

\title{\boldmath Dirac Sigma Models from Gauging}

\author[a]{Vladimir Salnikov}
\author[b]{Thomas Strobl}

\affiliation[a]{Laboratoire de Math\'ematiques de l'INSA de Rouen \\
Avenue de l'Universit\'e, 76801 Saint-\'Etienne-du-Rouvray cedex, France}
\affiliation[b]{Institut Camille Jordan,
Universit\'e Claude Bernard Lyon 1 \\
43 boulevard du 11 novembre 1918, 69622 Villeurbanne cedex,
France}

\emailAdd{vladimir.salnikov@insa-rouen.fr}
\emailAdd{strobl@math.univ-lyon1.fr}

\abstract{The G/G WZW model results from the WZW-model by a standard procedure of gauging. G/G WZW models are members of Dirac sigma models, which also contain twisted Poisson sigma models as other examples. We show how the general class of Dirac sigma models can be obtained from a gauging procedure adapted to Lie algebroids in the form of an equivariantly closed extension. 
The rigid gauge groups are generically infinite dimensional and a standard gauging procedure 
would give a likewise infinite number of 1-form gauge fields;
the proposed construction yields the requested finite number of them.

Although physics terminology is used, the presentation is kept accessible also for 
a mathematical audience.

\vspace*{1.0em}
}

\begin{document} 
\maketitle
\flushbottom

\section{Introduction and Motivation}
Gauging has always been an important technique in the construction of physical theories. For a very simple example of this procedure, consider the action functional of two free scalar fields $\varphi_1$, $\varphi_2 \in C^\infty(\S)$ over Minkowski space $\S$: Combining them into one complex scalar field, $\phi = \varphi_1 + i \varphi_2$, and adding a (rotation-invariant) potential of interactions $V$, the functional takes the form 
\be \label{eq:baby}
S[\phi] := \int\limits_\S  \partial_\mu \phi \, \overline{\partial^\mu \phi} + V(|\phi|^2) \: \rd^4x \, ,
\ee
where the bar denotes complex conjugation. Clearly $S$ is invariant under internal (rigid) rotations, which in terms of the complex field become phase transformations,
\be \phi(x) \mapsto e^{i\alpha}\phi(x) \, ,  \label{eq:phase}\ee
and it is not \emph{gauge invariant} under those symmetries, i.e.~the functional is not invariant if $\alpha$ is permitted to change together with $x$, $\alpha=\a(x)$. One can fix this ``deficiency'' by introducing a gauge field $A_\mu(x)$, replacing any derivative by a covariant one, $\partial_\mu \to \rD_\m = \partial_\mu - iA_\m$. The functional, now depending on $\phi$ and $A_\mu$, becomes gauge invariant by adding the transformation law 
 $A_\m \mapsto A_\m + \partial_\m \a$ for $A_\m$ to the (local) phase transformation of the form (\ref{eq:phase}) for $\phi$. 
 
In a physically more realistic setting one would start with free fermions instead of scalar fields, but the procedure is essentially the same. The obtained $A_\mu$ describes the photon, while, by increasing the number of initial (fermionic) fields, one constructs the Standard Model of elementary particle physics in this way, with the additional gauge fields describing now also the W- and Z-bosons  as well as the gluons. 

This procedure, called minimal coupling, works even in the much more general context given by sigma models.  Let $(\S,h)$ be any (pseudo)Riemannian $d$-dimensional manifold and replace the ``internal space'' $\R^2$ above, in which the rigid rotations took place, by any Riemannian 
manifold $(M,g)$ where the metric $g$ has a nontrivial isometry group $G$. 
Infinitesimally, the condition on $g$ reads 
\be \CL_v g = 0 \, , \label{eq:Kill}
\ee
valid for the vector fields  $v=\r(\xi)$ on $M$ corresponding to arbitrary elements  $\xi\in \g \equiv \Lieg$, $\r$ denoting the action of $\g$ on $M$ induced by the $G$-action. Consider the functional of maps $X \colon \Sigma \to M$ 
  \be \label{eq:sigma}
S[X] = \int\limits_\Sigma  \frac{1}{2} g_{ij}(X) \, \md X^i \wedge * \md X^j 
+  \int\limits_{\S} X^* B ,
\ee
where in the first term, which generalizes the first term of (\ref{eq:baby}) to $n=\dim M$ scalar fields taking values in the potentially curved space $M$, the star denotes the Hodge duality of differential forms on $\S$ induced by $h$. Instead of the potential term in (\ref{eq:baby}), which we could consider in principle as well, we put the pullback of a $d$-form $B$ defined on $M$. $G$-invariance of the potential is now replaced by $G$-invariance of $B$; in fact, this invariance needs to hold up to a ``total divergence'' only: Assuming the existence of a $(d-1)$-form $\b$ for any $v$ like in (\ref{eq:Kill}) such that 
\be \CL_v B = \rd \b \, , \label{eq:alpha}
\ee
the action (\ref{eq:sigma}) becomes invariant under the rigid symmetry group $G$. 

In the case $\beta=0$, the gauging of such a sigma model is again provided by minimal coupling. The gauge fields to be introduced are collected into a Lie algebra valued 1-form on $\S$ corresponding to a connection in a trivial principal bundle $\Sigma \times G$, $A\equiv A^a e_a \in \Omega^1(\S,\g)$, $e_a$ denoting a basis of $\g$. Again we merely need to replace $\rd X^i$ everywhere by the covariant derivative $\rD_A X^i \equiv \rd X^i - \rho^i_a(X) \, A^a$, where $i=1, \ldots , n$ and $\rho(e_a)\equiv \rho_a^i(X)\partial_i$ for $a = 1, \ldots,  \dim \g$. While $S[X]$ is invariant w.r.t.~the group $G$, the extended functional of $X$ and $A$  is invariant w.r.t.~the much bigger group $\CG$ of smooth maps from $\S$ to $G$.\footnote{$G$ is called a \emph{rigid} symmetry group and also coincides with the structure group of the bundle, $\CG$ is the \emph{gauge} (or \emph{local}) symmetry group.} Since any group of isometries (\ref{eq:Kill}) is finite dimensional, here necessaril
 y $\dim G< \infty$, while certainly $\dim \CG= \infty$. 
Below we will consider a context where already the rigid symmetry group $G$ is infinite dimensional (and still not to be confused with the gauge group $\CG=C^\infty(\S,G)$).

The situation becomes more complicated when $\beta \neq 0$ or when one even generalizes the second term in the sigma model to a Wess-Zumino term. Let $\tilde{\S}$ be a $(d+1)$-dimensional manifold with boundary $\S$, $\tilde{X} \colon \tilde{\S} \to M$ restricting to $X$ on the boundary, 
$H$ a closed $(d+1)$-form on $M$, and replace $\int_\S X^* B$ by 
\be S_{W\!Z} = \int\limits_{\tilde{\S}}  \tilde{X}^* H . \label{eq:WZ}\ee
Clearly, by Stokes' theorem, this reduces to the original term if $H=\rd B$.\footnote{For non-exact closed forms $H$, the functional is multi-valued as a functional of $X$. One may assume the conditions to be satisfied such that its contribution to the path integral is unique (cf.~the discussion in \cite{Witten}), but in \emph{any} case the \emph{variation} of $S_{W\!Z}$ depends on $X$ and its variation only.} Using Cartan's magic formula, $\CL_v = \rd \iota_v + \iota_v \rd$, the condition (\ref{eq:alpha}) on $B$ can be rewritten as
\be \iota_v H = \rd \a \, , \label{eq:theta}
\ee 
for $H=\rd B$ (here we have put $\a = \b- \iota_v B$) and it is this condition one requires for a general, also non-exact $H$. One sees at once that now minimal coupling would not provide a satisfactory result: Replacing $\rd X^i$ by $\rD_A X^i$, would contain a term with the 1-forms $A^a$ taken to the $(d+1)$-fold wedge-power  and even its variation would not localize to a term defined on the boundary $\S$. In fact, in general there exist \emph{no} additions to the action functional (that are local on $\S$) such that the sigma model with WZ-term becomes gauge invariant, in which case one speaks of an ``anomaly''. There is no anomaly, iff the given $(d+1)$-form $H$ permits an equivariantly closed extension \cite{Stanciu,etc}. Alternatively, there is no anomaly, iff the couples $(v,\a)$ in (\ref{eq:theta}) used for gauging form a Dirac structure \citep{Alekseev-Strobl}. We will come back to this below. 

An important example of this is provided by the WZW-model \citep{Witten,??}. This is a two-dimensional sigma model or string theory, $\dim \S =2$, with the target space $M$ being chosen to be a semi-simple Lie group. The metric $g$, used for the kinetic term in (\ref{eq:sigma}), is the, e.g., left-invariant extension of the Killing-metric, the closed 3-form $H$ is given by the Cartan-Killing 3-form; in a faithful matrix representation of the group, 
and up to an irrelevant prefactor which we fix by some convention, $H=\frac{1}{3} tr (\gr^{-1} \rd \gr)^{ \wedge 3}$. This model has a rigid symmetry group that is given by two copies of the target group, corresponding to left- and (independent) right-translations. Gauging this large rigid symmetry is obstructed, only particular subgroups permit gauging, like choosing $G$ equal to the target group itself (with the adjoint action); gauging then yields the so-called G/G WZW model \citep{???!!!}:
\ba S_{G/G}[\gr,A] = S_{kin}[\gr,A] + \nonumber \\ 
 \int\limits_\S  \tr\left(A\wedge( \gr^{-1}\wedge \rd \gr - \rd \gr \wedge \gr^{-1}) + A \wedge \gr  A  \gr^{-1} \right)
+ S_{WZ} \label{eq:G/G}
\ea 
where $S_{kin}$ is found to be just the usual kinetic term with minimal coupling, $S_{kin} = \frac{\lambda}{2} 
\int_\S tr (\gr^{-1} \rD_A \gr \wedge *  \gr^{-1} \rD_A \gr)$ with $\lambda=1$ (the real auxiliary parameter $\lambda$ is introduced for later convenience). 
Using even smaller subgroups leads to $G/H$ WZW models \citep{Gawedzki}. While these coset G/H models still carry ``physical'' degrees of freedom and describe interesting string theories, the $G/G$ WZW model becomes topological \cite{Gerasimov}. 

There is another important topological field theory in two dimensions, the Poisson sigma model (PSM) \citep{Ikeda, Schaller-Strobl}, which was used by Kontsevich to find his famous quantization formula \citep{Kontsevich} (cf.~also \citep{Cattaneo-Felder}), but which also permits to concisely treat a large class of two-dimensional gravity-Yang-Mills theories \cite{Kloesch-StroblI, Kloesch-StroblII,Kloesch-StroblIII}.  It was generalized to carry a WZ-term in \cite{Klimcik-Strobl}, in which case it takes the form
\be \label{eq:HPSM}
S[X,A]= \int_\Sigma  A_i \wedge \md X^i + \frac{1}{2}
\Pi^{ij}(X) A_i \wedge A_j + \int_{\tilde{\S}} \tilde{X}^*H ,
\ee
where in addition to $n=\dim M$ scalar fields $X^i$ the action depends also on a likewise number of 1-form fields $A_i$. $\Pi\equiv \Pi^{ij} \partial_i \wedge \partial_j$ is a bivector field on $M$. For $H=0$, i.e.~in the conventional PSM, $\Pi$ is Poisson. The model (\ref{eq:HPSM}), on the other hand, is topological, iff \citep{Klimcik-Strobl} 
the couple $(\Pi,H)$ satisfies the twisted Poisson condition $ \frac{1}{2} [\Pi, \Pi] = \langle H,\Pi^{\otimes 3} \rangle$  (cf.~also \cite{Park,Severa-Weinstein}). 

A joint generalization of the twisted Poisson sigma model (\ref{eq:HPSM}) and the G/G-WZW model is given by the so-called Dirac sigma model (DSM) \citep{Kotov-Schaller-Strobl}. In fact, like the G/G-model (\ref{eq:G/G}), also the DSM carries a kinetic term (with minimal coupling to $A$). In both of these two models there are various arguments \citep{Gerasimov,Kotov-Schaller-Strobl}, moreover, that the physical content of the theory does not change when dropping this term, i.e.~taking the limit $\lambda \to 0$. For the special case of a Dirac structure being a twisted Poisson structure, after this limit, the action of the DSM reduces to (\ref{eq:HPSM}), while for the Dirac structure corresponding to the adjoint action on a group we are left likewisely with the last three terms in (\ref{eq:G/G}).\footnote{For the general DSM, the resulting action takes the form (\ref{eq:DSM}) or (\ref{eq:DSMtop}); the notions needed to understand the terms in that functional are explained in the course of the present paper or in the original article \cite{Kotov-Schaller-Strobl}.} The structural similarity of these terms is striking. But while in the first example the $A$-contributions (which do not result from a simple procedure like minimal coupling) can be obtained from a gauging procedure  
as outlined above, namely by an equivariantly closed extension of the 3-form $H$, the much more general Dirac sigma model, or even its special case of the twisted PSM (\ref{eq:HPSM}), was never \emph{derived} or explained like this. 

It will turn out that the groups $G$ coming into question for the gauging will all be infinite dimensional.\footnote{Note in this context that for what follows we dropped the kinetic term and thus the metric $g$, which, if required to be left invariant, would necessarily lead to a finite dimensional $G$. The reason for this is at least two-fold here: First of all, for kinetic terms the simple recipe of minimal coupling always works. Second, as argued above, in the resulting theories the physics does not change, if the kinetic term is dropped (limit $\lambda \to 0$).} Nevertheless, as we see from the answer like (\ref{eq:HPSM}) we want to reproduce, the number of gauge fields is finite. This is an interesting fact for several reasons: First, the usual equivariant cohomology procedure yields precisely the same number of 1-form gauge fields as the dimension of the group $G$. 
So, the generalized equivariant cohomology procedure using an adequate BRST-type language that we will sketch below, is different from the usual one and capable of producing interesting results. Second, and more important, in the models under consideration here, 
one deals with Lie groupoids/algebroids as generalizations of Lie groups/algebras. 
These are, on the one hand, generically infinite dimensional Lie groups/algebras, on the other hand, they are themselves finite dimensional manifolds (cf., e.g., \citep{Weinstein,...} for an overview). But they are also much more flexible than mere groups, which, in this context, appear as somewhat (too?) rigid: in the space of all Dirac sigma models the G/G models appear as isolated points (such as linear Poisson brackets in the world of arbitrary Poisson structures). So, developping formalisms which permit such a type of symmetries for the construction of new theories is a promising direction, all the more if the resulting theories \emph{do} resemble the more traditional ones quite closely 
(like in the comparison of (\ref{eq:G/G}) and (\ref{eq:HPSM}) of our toy models). Certainly such a formalism will not be restricted to two spacetime dimensions, we do it here only in a first step for the purpose of developping the new formalism. We remark in this context, that another step into the direction of developping (also theories of potentially physical content and in arbitrary dimensions of $\S$) was made in \citep{LAYM,Mayer-Strobl,GeneralizingGeometry}. We consider the present investigation as complementary to the tools developped there.

\section{Dirac structures/Lie algebroids and the group G to gauge \label{sec:G}}
The goal we want to pursue now is gauging the WZ-term (\ref{eq:WZ}) for a given closed 3-form $H$. For this purpose we need to first identify the group $G$ of rigid symmetries that can be gauged. We already found above, having a \emph{symmetry} of $H$ generated by a vector field $v\in 
\Gamma(TM)$ should be read as the existence of a corresponding 1-form $\a\in 
\Gamma(T^*M)$ such that (\ref{eq:theta}) holds true. Now the question of obstructions in the gauging arises. 

\begin{figure}[b] \centering
\includegraphics[width=0.5\linewidth]{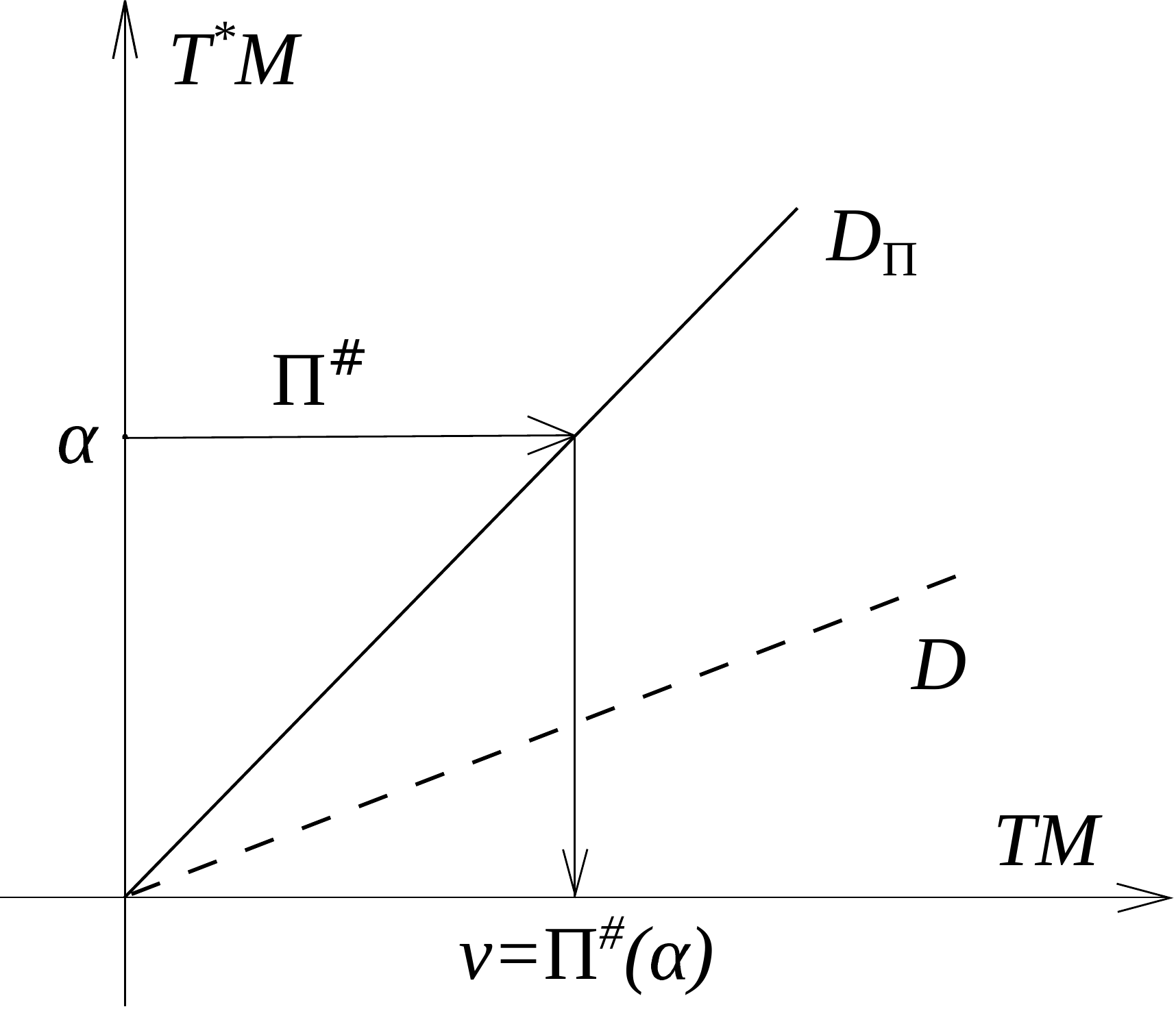}
\caption{\label{fig:Dirac}  Schematic picture of a Dirac structure $D_\Pi$ corresponding to a 
bivector $\Pi$ and one that does not arise in this way (having some directions parallel to $TM$).}
\end{figure}

Here we take recourse to \cite{Alekseev-Strobl}: It was found that, for a very large class of two-dimensional sigma models, couples $(v,\a) \in \Gamma(TM \oplus T^*M)$ generate anomaly-free symmetries, iff they fit into a so-called Dirac structure. A Dirac structure is a maximal collection of such couples, such that for any two of them, $(v,\a)$ and $(w,\b)$, one has: 1. $\a(w) + \b(v)=0$ and 2. The Courant-Dorfman bracket of the two,
\be \label{eq:Dorfmann}
[(v,\a),(w,\b)]_{D} = ([v,w], \CL_v \b - \iota_w \rd \a + \iota_{w} \iota_v H)
\ee 
gives another one in the set of permitted couples. We remark that taking the first condition into account, the bracket becomes a Lie bracket. An important example  \cite{Severa-Weinstein} of a Dirac structure is provided by $(\Pi,H)$ defining an $H$-twisted Poisson structure: In this case one takes all couples $(\iota_\a \Pi, \a)$, 
parametrized by 1-forms $\a \in \Omega^1(M)$. In fact, given any  tensor field $\Pi\in \Gamma(TM^{\otimes 2})$ one can define a subbundle $D_\Pi \subset TM \oplus T^*M$ by means of the graph of the map  $\Pi^\sharp \colon T^*M \to TM, \a \mapsto \iota_\a \Pi \equiv \Pi(\a,\cdot)$. It turns out that $D_\Pi = \{ (\Pi^\sharp(\a), \a), \a \in T^*M \}$ is a Dirac structure, iff $\Pi$ is $H$-twisted Poisson \cite{Severa-Weinstein}. But not any Dirac structure $D$ arises in this way; there can be directions of $D$ parallel to some directions of $TM$ (e.g.~for $H=0$, $D=TM$ is a Dirac structure). Figure \ref{fig:Dirac} gives a schematic overview.

Now we are ready to define the biggest possible unobstructed rigid symmetry Lie algebras $\g$ of a given closed 3-form $H$. To avoid anomalies we need to choose a Dirac structure $D \subset TM \oplus T^*M$. Then 
\be \g = \{ (v,\a)\in \Gamma(D) | \iota_v H = \rd \a \} \, . \label{eq:g}
\ee 
Several remarks are in place: The set of all elements $(v,\a)$ satisfying (\ref{eq:theta}) forms an algebra with respect to the bracket (\ref{eq:Dorfmann}); this algebra is a Leibniz algebra, but not Lie due to the missing antisymmetry of that bracket. Restricted to sections $\Gamma(D)$ of any Dirac structure $D$, however, the bracket becomes a Lie bracket. While, on the other hand, the Lie algebra $\Gamma(D)$ forms a $C^\infty(M)$-module, which implies that it arises from a subbundle $D$ of $TM\oplus T^*M$, the condition (\ref{eq:theta}) does not: If $(v,\a)$ satisfies the condition (\ref{eq:theta}), then, for a general function $f\in C^\infty(M)$, $(fv,f\a)$ does not. Correspondingly, while any Dirac structure $D$ defines a particular Lie algebroid structure over $M$, the Lie algebra $\g$ does not even have the interpretation of sections of any (sub)vector bundle. Note also that generically the Lie algebra $\g$ will be infinite dimensional. Certainly we can still decide to 
 look for (also finite dimensional)
 subalgebras $\h \subset \g$ to be gauged, e.g.~by requiring that $\h$ leaves invariant also some given metric $g$ on $M$. We will not restrict us here like this for what follows, but take any group $G$ the Lie algebra of which is given by $\g$. 

Two final remarks: It is curious to note that the condition (\ref{eq:theta}) reminds of the condition for $v$ to be a Hamiltonian vector field. In fact, if $H$ is a non-degenerate closed $(d+1)$-form, it can be viewed of as a higher analogue of a symplectic form and such vector fields $v$ are then called $d$-Hamiltonian \cite{Rogers}. Finally, one may ask, why the consideration in \cite{Alekseev-Strobl} can be applied since it uses the cotangent bundle of loop space although the kinetic term for the string coordinate $X$ was dropped. In fact, whenever the gauging of a WZ-term (\ref{eq:WZ}) will produce a term linear in the gauge field (like in (\ref{eq:G/G}) or also in (\ref{eq:HPSM})), the spatial component of the gauge field $A$ (possibly multiplied by an appropriate function of $X$) will provide a momentum $p$ conjugate to $X$.\footnote{For a more detailed and careful discussion of this point, one may refer to the last section of \cite{Kotov-Schaller-Strobl}, where it is 
 argued in what sense one can always work with the same phase space, the kinetic term being present or not.}

\section{BRST-picture of standard and Lie algebroid equivariant cohomology}\label{sec:BRST}
To gauge the Lie algebra $\g$ in (\ref{eq:g}) as a rigid symmetry of the WZ-term (\ref{eq:WZ}), we will apply the generalization of the standard equivariant cohomology as developped in \cite{Kotov-Strobl}. For this purpose we first recall the standard one, defined by a finite dimensional Lie algebra $\h$ (which may be the intersection of $\g$ with the isometry group of a metric $g$ defined on $M$, or just an arbitrary finite dimensional Lie algebra acting on $M$). We will present the ordinary gauging such that the generalization to $\g$ will be straightforward (cf.~also \cite{Kalkman}).

Denote by $\rho \colon \h \to \mathfrak{X}(M), e_a \mapsto \rho^i_a \partial_i$ the action of $\h$ on $M$, where $(e_a)_{a=1}^{\dim \h}$ is a basis of $\h$ and $\rho^i_a \in C^\infty(M)$. Then there is a canonical BRST-charge associated to it \cite{?}:
\be Q = \xi^a \rho_a^i(x) \frac{\partial}{\partial x^i} - \frac{1}{2} C^a_{bc} \xi^b\xi^c
 \frac{\partial}{\partial \xi^a} \, , \label{eq:Q}
\ee 
where $x^i$ are coordinates on $M$, $\xi^a$ are odd, degree one variables called ``ghosts'' in the BRST language, and $C^a_{bc}$ are the structure constants of the Lie algebra. Let $E$ be the trivial vector bundle $M \times \h$. Declaring fiber-linear coordinates on $E$ to have degree one (and thus to be odd), we can consider $x^i$ and $\xi^a$ as coordinates on this graded super-manifold $\CM$ that is usually denoted by $E[1]$. $Q$ is a vector field of degree +1 on $\CM$ that squares to zero by the BRST property. 

Given more generally any vector bundle $E$ over $M$, we can introduce such coordinates on $E[1]$. A degree +1 vector field then always takes the form  
(\ref{eq:Q}), where now, however, also $C^a_{bc}$ can depend on $x$. Requiring $Q^2=0$ then can be used as a possible definition of the structure of a Lie algebroid \cite{Vaintrob}.\footnote{According to a more conventional definition, a Lie algebroid is a vector bundle $E\to M$ equipped with a Lie bracket $[ \cdot , \cdot ]_E$ on its sections $\psi$, $\psi'$ and a bundle map $\rho \colon E \to TM$ 
such that for any function $f \in C^\infty(M)$ one has $[\psi, f \psi']_E = f [\psi,  \psi']_E + \left(\rho(\psi) \cdot f\right) \psi'$. The relation with $Q$ 
is as follows: For any, possibly overcomplete basis $e_a \in \Gamma(E)$ of sections in the bundle, one has $[e_a, e_b]_E|_x = C^a_{bc}(x)$ and $\rho(e_a)|_x = \rho_a^i(x) \partial_i$ for all $x \in M$. Alternatively one can identify the 
Lie bracket $[ \cdot , \cdot ]_E$ with the bracket defined in (\ref{eq:der}) 
below.}
Dirac structures are particular Lie algebroids (cf., e.g., \cite{Kotov-Schaller-Strobl}) and thus they also permit the formulation in terms of a homological $Q$, of which we will make use below in the generalization. Lie algebroids with $C^a_{bc}$ constant always come from a Lie algebra action and are called action Lie algebroids. 

For any Lie algebroid, the degree -1 vector fields $\epsilon, \,\epsilon'$ 
form a Lie algebra with respect to the derived bracket \cite{Kosmann-Schwarzbach}:
\begin{equation}\label{eq:der}
[\epsilon , \epsilon']_Q:=[[\epsilon , Q ] , \epsilon'] \, , 
\end{equation}
where the ordinary brackets denote graded commutators. In the case of (\ref{eq:Q}) and the action Lie algebroid, this reproduces the usual Lie bracket in $\h$ for the case of constant (w.r.t.~the natural flat connection on the trivial bundle) vector fields, $\epsilon = \epsilon^a \frac{\partial}{\partial \xi^a}$ 
 with $\epsilon^a$ constant. Note, however, that also in this case $\epsilon^a \in C^\infty(M)$ is permitted and such sections form an infinite dimensional Lie algebra (which for $\rho=0$ would be just $C^\infty(M) \otimes \h$, but in general includes also a differentiation w.r.t.~the base manifold due to the first term in (\ref{eq:Q})). 

Suppose now that $H$ is a closed 3-form satisfying (\ref{eq:theta}) for any vector field $v$ originating from the $\h$-action. So, for any $v_a := \rho(e_a)$ we have an $\alpha_a$ such that $H(v_a,\cdot,\cdot)=\rd \a_a$. Note that, even if they exist, these $\alpha_a$ are not unique; they will have to be chosen in such a way so as to satisfy some additional requirements. All this, including even the condition  (\ref{eq:theta}) in fact, will follow from the procedure below. 

Being a 3-form on $M$, $H$ can be also regarded as a 3-form on the bundle $E$ over $M$ (when pulled back by the projection, denoted by the same letter for simplicity), and thus likewise as a 3-form on the graded manifold $\CM$, $H \in \Omega^3(\CM)$. Let us define a total degree $\deg$ by adding the form degree and the ghost degree. Since $H$ does not contain any $\xi^a$-variables, $H$ has also total degree 3. 

We noticed above that the $\h$-action on $M$ can be generated by (constant) degree -1 vector fields $\epsilon$ on $\CM\equiv M \times \h[1]$; we have in particular, $({\mathfrak{X}}_{-1}^{const}(\CM), [ \cdot , \cdot ]_Q)\cong (\h,[ \cdot , \cdot ])$. An $\h$-equivariantly closed extension $\widetilde{H}$ of $H$ can now be defined as follows \cite{Kotov-Strobl}: $\widetilde{H}$ is a differential form on $\CM$ of total degree 3, $\deg \widetilde{H}=3$, such that 
\begin{eqnarray} 
  \widetilde{H}|_{\xi^a=0=\rd \xi^a}=H  \; , \qquad (\rd + \CL_Q) \widetilde{H} = 0 \; , \qquad
  \CL_\epsilon \widetilde{H} = 0\; , \label{eq:equiv}
\end{eqnarray}
where the last equation has to hold for all $\epsilon \in {\mathfrak{X}}_{-1}^{const}(\CM)$. The first equation ensures that at ghost number zero $\widetilde{H}$ starts with the given 3-form $H$ (that is why it is called an extension), the last equation translates into the fact that $\widetilde{H}$ depends on the ghosts only through their differentials; since $\deg(\rd \xi^a)=2$, there can be at most one of them, so $\widetilde{H}=H+ \alpha_a \wedge \rd \xi^a$ for some 1-forms $\alpha_a$ on $M$. It remains to solve the second condition of (\ref{eq:equiv})
at each ghost number. At ghost number 0 and 1 it is fulfilled by construction, saying that $H$ is closed and reproducing the condition (\ref{eq:theta}) valid for each couple $(v_a,\alpha_a)$, $v_a \equiv \rho_a^i \partial_i$, respectively, i.e., in the second case, reducing to the rigid invariance condition of $H$ w.r.t.~the $\h$-action on $M$.\footnote{We already adapted the notation so as to identify the 1-form part of $\tH$ with the $\alpha_a$ appearing in the condition (\ref{eq:theta}), which otherwise follows from the degree 1 equation (note that in condition (\ref{eq:theta}) $\alpha_a$ is defined only up to closed contributions).} The remaining two equations at ghost numbers 2 and 3 provide the constraints that cannot always be fulfilled: $\CL_{v_a} \alpha_b = C^c_{ab} \alpha_c$ and $\iota_{v_a} \alpha_b + \iota_{v_b} \alpha_a=0$. 
Note that the first of these conditions would be the equivariance condition of a moment map if $\alpha_a$ were a function on a symplectic manifold; it thus can be interpreted as the corresponding higher analogue for the case that $(M,H)$ is 2-symplectic. This reproduces the standard formulas for an equivariant extension, cf., e.g., \cite{Stanciu}, while it is formulated in such a way that a generalization to arbitrary Lie (and also higher Lie or Lie-n) algebroids will be straightforward.

But before turning to this, we will briefly describe how to obtain the gauge invariant functional from such a procedure, i.e.~after having found $\widetilde{H}$. The scalar fields $X^i$ we started with can be viewed as the pullback of coordinate functions $x^i$ on $M$ by the map $X \colon \Sigma \to M$, $X^i = X^*(x^i)$. The original functional $S$ depends on the map $X$ only. To include also the gauge fields, for each dimension of $\h$ one 1-form gauge field on $\Sigma$, we may extend the above map $X$ to a degree-preserving map $a \colon T[1]\Sigma \to E[1]\equiv M \times \h[1]$. Since coordinates on $T[1]\Sigma$ are coordinates $\sigma^\mu$ on $\Sigma$ together with $\rd \sigma^\mu$, and the latter coordinates are declared to have degree 1 and thus being odd, functions on $T[1]\Sigma$, i.e.~of these two type of variables, are nothing but differential forms on $\Sigma$. Thus the pullback of the degree 1 variable $\xi^a$ on $\h[1]$ has to be a 1-form on $\Sigma$. We thus 
 have the identification: $X^i = a^*(x^i)$, $A^a =a^*(\xi^a)$. 

$\widetilde{H}$ is a differential form on $\CM=E[1]$ and 
can thus be viewed as a function on $T[1]\CM$ by the above consideration, now extended to the case of graded manifolds (instead of the ordinary manifold $\Sigma$). So, if $q^\alpha\equiv (x^i,\xi^\alpha)$ denote the graded coordinates on $\CM$, we need to extend the map $a$ further to a map $f \colon T[1]\Sigma \to T[1]\CM$ by prescribing their action by pullback on the $\rd q^\alpha$s. Noting that $\rd q^\alpha = \rd (q^\alpha)$ and that the de Rham differential is a vector field on $T[1]\CM$ as well as on $T[1]\Sigma$ (we denote it by the same letter $\rd$, understanding their difference from the context), the simplest way of fixing $f^*$ would be to say that it commutes with $\rd$. Let us denote such a map by $f_0$. Then $(f_0)^*(q^\alpha)=a^*(q^\alpha)$ and $(f_0)^*(\rd q^\alpha)=\rd a^*(q^\alpha)$; so, for example, $(f_0)^*(\rd \xi^a)=\rd A^a$. It is, however, useful to twist this map by a diffeomorphism generated by $\iota_Q \equiv Q^\alpha(q) \frac{\partial}{\partial \
 \rd q^\alpha}$,
\begin{equation}
\exp(\iota_Q) (q^\alpha) = q^\alpha \; , \quad \exp(\iota_Q) (\rd q^\alpha) =\rd q^\alpha+ Q^\alpha(q) \; ,
\end{equation} 
i.e.~shifting the image along the tangent direction by the value of the vector field $Q$:
\begin{equation}
f^* = \exp(\iota_Q) \circ (f_0)^* \circ \exp(-\iota_Q)\: .
\end{equation}
Then, by construction, $f^* \colon C^\infty(T[1]\CM) \to C^\infty(T[1]\Sigma)$ 
is a chain map with respect to the twisted de Rham differential $\exp(\iota_Q) \circ \rd \circ \exp(-\iota_Q) = \rd + \CL_Q =: \widetilde{Q}$ on $\CM$, i.e.~$\rd \circ f^* = f^* \circ  \widetilde{Q}$. For the vector field $Q$ of the action Lie algebroid (\ref{eq:Q}), this map $f$ has the remarkable property of implementing minimal coupling and producing the curvature of the gauge field:
\begin{equation}
f^*(\rd x^i) = \rd X^i - \rho^i_a(X) A^a \equiv D_A X^i \; , \quad f^*(\rd \xi^a) = \rd A^a + \frac{1}{2} C^a_{bc} A^b \wedge A^c \equiv F_A \, ,
\end{equation}
while it still produces the scalar and gauge fields when applied to the coordinates on $M\times \h[1]$: $f^*(x^i) = X^i$ and $f^*(\xi^a) = A^a$. The recipe\footnote{For an explanation cf, e.g., \cite{Stanciu} or \cite{Kotov-Strobl}.} now goes as follows: Let $S[X]=\int_\Sigma X^*B$ be invariant under an $\h$-action, or (\ref{eq:WZ}) its corresponding Wess-Zumino generalization, and let $\widetilde{H}$ be an equivariant extension of $H$ as defined above (in the first case, $H=\rd B$), then the gauge invariant extension of this action\footnote{In the WZ-generalization one would need to add ``variation'' in front of ``action''.} is a functional of $a$ and takes the form
\begin{equation} \label{eq:action}
S[a] \equiv S[X,A] = \int_{\widetilde{\Sigma}} \widetilde{f}^* (\tH) \; , \qquad \partial \widetilde{\Sigma} =\Sigma \, ,
\end{equation}
where on the rhs $\widetilde{f}$ is the map $f$ for the extension of $a$ to $\widetilde a \colon \colon T[1] \widetilde{\Sigma} \to \CM$. Note that $\tH$ has total degree 3 so that we obtain a 3-form on $\widetilde{\Sigma}$ by applying the degree preserving map $\widetilde{f}^*$  to it; except for the WZ-term, this 3-form turns out to be exact and thus the integral over the additional terms indeed localizes to the boundary $\Sigma$. 

Now the stage is set for the Lie algebroid generalization of the equivariant cohomology. Let $H$ be a closed 3-form on $M$ and $\g$ as defined in (\ref{eq:g}). Then $\tH$ is called an equivariantly closed extension of $H$ in the Lie algebroid sense, or, for short, an $E$-equivariantly closed extension of $H$ where $E$ is the Lie algebroid under consideration, if
\begin{eqnarray} 
  \widetilde{H}|_{\mathrm{gh} 0}=H  \; , \qquad \widetilde{Q} \widetilde{H} = 0 \; , \qquad
\widetilde{\epsilon}   \widetilde{H} = 0\; . \label{eq:Eequiv}
\end{eqnarray}
Here the first equation denotes setting to zero all coordinates of strictly positive degree on $E[1]$ as well as their derivatives. 
$\widetilde{Q}=\rd + \CL_Q$ is the twisted de Rham differential. 

More intricate is the specification of the third equation. In any case,  $\widetilde{\epsilon}$ has to form a subalgebra of $({\mathfrak{X}}_{-1}(T[1](E[1])),
[\cdot,\cdot]_{\tQ} \equiv [[\cdot, \tQ],\cdot]$, which is Lie. One may demand in addition that $\widetilde{\epsilon} = \CL_\epsilon$ where $\epsilon \in \mathfrak{X}_{-1}(E[1])$, which can be  identified with the sections of the Lie algebroid $E$. If $E=M \times \h$ is an action Lie algebroid, we will require that one considers the sub-Lie algebra of constant sections on $E$ corresponding to elements of $\h$; this then reproduces standard equivariant cohomology. For the case of a Dirac structure, $E=D$, we will require it to satisfy the condition corresponding to the definition of $\g$ in (\ref{eq:g}), or one of its Lie subalgebras. However, we will find it necessary or at least useful below, to ``lift'' this picture, i.e.~to permit vector fields $\te$ on $T[1]E[1]$ that do not come from $E[1]$ in the above way; in that case, they will correspond to an extension $\widetilde{\g}$ of the Lie algebra $\g$.

Note that applying the standard equivariant extension with $\h=\g$ in general leads to a different functional than applying the $D$-equivariant one for the same $\g$; this is particularly transparent when $\g$ is infinite dimensional so that the former procedure will introduce an infinite number of gauge fields, while the one corresponding to the Dirac structure will introduce a finite number of them only, at the price of them being of a Lie algebroid type instead of a conventional gauge field. We will comment more on these two alternatives and their relation in a separate paper \cite{Salnikov-Strobl2}.

\section{The Dirac sigma model from Gauging}

As a warm-up we start with the Poisson sigma model, i.e.~with (\ref{eq:HPSM}) for $H=0$ and $\Pi$ being a Poisson bivector. We thus first determine the maximal Lie algebra $\g$ to gauge for this special case. Since $H=0$, the 1-forms $\alpha$ have to be closed, $\rd \alpha = 0$. Moreover, $v$ is completely determined by $\alpha$, $v = \Pi^\sharp(\alpha)$ (this becomes also clear from Fig.~1, where we see that the Dirac structure $D_\Pi$ can be identified with $T^*M$). Let us assume for a moment that $H^1(M)=0$, i.e.~that closed 1-forms are already exact, $\alpha = \rd f$. In this case the bracket (\ref{eq:Dorfmann}) reduces to simply $[ \rd f, \rd g]_D= \rd \{f,g\}_\Pi$ where
$\{\cdot,\cdot \}_\Pi$ is the Poisson bracket on $M$ induced by $\Pi$, $\{f,g\}_\Pi=\iota_{\rd g}\iota_{\rd f} \Pi \equiv \left(\Pi^\sharp(\rd f)\right)
(g)$. Thus, for  $H^1(M)=0$ the Lie algebra $\g$ is isomorphic to the Poisson algebra on $M$ modulo constants, $(\g,[ \cdot, \cdot]) \cong (C^\infty(M)/\mathrm{const},\{ \cdot , \cdot \}_\Pi)$. In general, $\g$ can be identified with $(\Omega^1_{\mathrm{closed}}(M), [ \cdot ,\cdot ]_D)$, which also fits into the following exact sequence of Lie algebras
\begin{equation}
0 \to (C^\infty(M)/\mathrm{const},\{ \cdot , \cdot \}_\Pi)\stackrel{\rd}{\to}
 \g\cong (\Omega^1_{\mathrm{closed}}(M), [ \cdot ,\cdot ]_D) \stackrel{[\:\cdot\:]}{\to} H^1(M)\to 0 \, ,
\end{equation}
where $H^1(M)$ is viewed upon as an abelian Lie algebra. This is true since the bracket between two closed 1-forms $\alpha$ and $\beta$ induced by the bracket (\ref{eq:Dorfmann}) is even exact, $\rd \left(\Pi(\alpha, \beta)\right)$, so vanishing when taking the cohomology class on the right.

Identifying $D_\Pi$ with $T^*M$ 
(cf.~Fig.~1), we may easily recover its Q-description. The graded manifold $\CM=T^*[1]M$ is symplectic, $\omega = \rd x^i \wedge \rd p_i$. It thus carries a canonical Poisson bracket $\{ \cdot , \cdot \}$ (of ghost number -1, since $\mathrm{gh}(\omega)=+1$). The bivector field $\Pi$ can be viewed as a Hamiltonian quadratic in the momenta, i.e.~of ghost number 2, $\Pi=\frac{1}{2} \Pi^{ij}(x)p_ip_j$. Thus its Hamiltonian vector field, $Q=\{\Pi, \cdot \}$ is a 
(ghost) degree +1 vector field. $\{\Pi,\Pi\}=0$ is equivalent to the Poisson condition on $\Pi$, so $Q$ squares to zero, $Q^2 \equiv Q \circ Q = 0$. Degree -1 vector fields $\epsilon = \alpha_i(x) \frac{\partial}{\partial p_i}$ are parametrized by 1-forms $\alpha = \alpha_i(x) \rd x^i$ on $M$. And indeed, the derived bracket (\ref{eq:der}) can be identified with the Lie algebroid bracket between the sections of $T^*M$ and agreeing with the Courant-Dorfman bracket (\ref{eq:Dorfmann}) restricted to sections of $D_{\Pi}$. Thus the Lie algebra $\g$ can be also seen as degree -1 vector fields $\epsilon$ on $\CM=T^*[1]M$ parametrized by $\alpha \in \Omega^1_{\mathrm{closed}}(M)$ and equipped with the derived bracket $[ \cdot , \cdot ]_Q$. 

Now we are ready to solve the gauging or extension problem (\ref{eq:Eequiv}) in this case, where we choose $\te=\CL_\epsilon$ with the degree -1 vector fields $\epsilon$ as above. First, we observe that the symplectic form $\omega$ on $\CM$ 
provides an obvious solution of this: It carries ghost number one, so it solves the first of the three conditions  (\ref{eq:Eequiv}). The second condition, $(\rd + \CL_Q) \omega=0$, is satisfied since the symplectic form is closed and $Q$ is Hamiltonian. It remains to check the last equation, the vanishing of $\CL_\epsilon \omega = \rd (\iota_\epsilon \omega) = - \rd (\alpha_i \rd x^i)\equiv - \rd \alpha$, which is the case due to the condition on the 1-forms $\alpha$. It is interesting to note that here it is the last equation that encodes the condition (\ref{eq:theta}) (here in the degenerate case $H=0$), in contrast to the standard equivariant cohomology, where it was found to show up in the second of these equations. This is a feature that persists also for non-vanishing $H$. 

Since we are extending $H=0$, certainly also $\tH= \lambda \omega$ for any $\lambda \in \R$ gives a solution. In fact, a direct, straightforward calculation shows that this is even the most general solution. We thus apply 
(\ref{eq:action}) to this solution:
\begin{equation}\label{eq:fomega}
\int_{\widetilde{\Sigma}} f^*(\lambda \omega) = \lambda \int_{\widetilde{\Sigma}} \rd \left(A_i \wedge \rd X^i + \frac{1}{2} \Pi^{ij}(X) A_i \wedge A_j\right) \, , 
\end{equation}
where we dropped a term proportional to $ \Pi^{ij}{},_l \Pi^{lk} \,A_i \wedge A_j \wedge A_k$ by means of the Jacobi identity satisfied by $\Pi$. Thus, up to an irrelevant constant prefactor, we indeed obtain the PSM on the boundary $\Sigma$. (Note that $(f_0)^*(\omega)=dX^i \wedge \rd A_i$, so again the twist by $Q$ is essential here.) For a generalization of (\ref{eq:fomega}) to arbitrary dimensions, where one obtains the AKSZ-sigma models on the boundary of the respective $\widetilde{\Sigma}$, cf.~\cite{Kotov-Strobl}.

It may appear strange to apply the extension problem to $H=0$. In that case all the information is contained in the chosen Dirac structure, which corresponded to the choice of a Poisson structure on $M$. To obtain from this a constant multiple of the Poisson sigma model in this way, and nothing else, is, however, comforting. It may be compared with applying standard equivariant cohomology to $M$ being just a point, where one obtains the cohomology of the chosen Lie algebra (and nothing else).\footnote{We are grateful to A.~Alekseev for this remark.}

From the initial example we already learn that one will have to impose some kind of non-zero condition on $H$ to hope for a uniqueness; otherwise, there could be factors of the above sort. But even requiring that $H$ is non-vanishing when restricted to the twisted symplectic leaves of  $\Pi$, the above procedure turns out to not uniquely fix the functional in general, as the following example shows.

The couple $(\Pi=\partial_1 \wedge \partial_2 + \partial_3 \wedge \partial_4+ x^1x^2\partial_2 \wedge \partial_3, 
H=-\rd\left(\frac{1}{2}(x^1)^2\rd x^2 \wedge \rd x^4\right))$ defines a twisted Poisson structure on $M=\R^4$. Since the bivector is non-degenerate, it is even twisted symplectic; thus there is only one twisted symplectic leaf, which is all of $\R^4$, and the 3-form is non-zero almost everywhere (it is non-vanishing except for the 3-plane $x^1=0$), ambiguities of the previous sort would thus be ruled out by a continuity argument. We next determine the Lie algebra $\g$ (\ref{eq:g}) for this choice of a 3-form $H$ and the given Dirac structure. Since the latter one is again the graph of a bivector, the condition to be in the Dirac structure is taken care of by using 1-forms $\alpha$ on $M$ to parametrize them. It remains to analyze the condition (\ref{eq:theta}), which now takes the form $\iota_{\Pi^\sharp(\alpha)} H = \rd \alpha$. The general solution of this is still an infinite dimensional space, but in some sense much smaller than in the previous example studied above: One 
 finds $\alpha =  g(x^1, x^4) \rd x^1 + 
 h(x^1,x^4) \rd x^4+c (x^1 x^2 \rd x^1 + \rd x^3) $, where $h$ is determined by $g$ up to an additive function of $x^4$ by means of $h_{,1} = -x_1 g + g_{,4}$. Thus while in the Poisson case studied before one obtained essentially all functions of the Poisson manifold parametrizing the Lie algebra, which would be functions of \emph{four} variables for $M=\R^4$, here the general solution is parametrized by one function $g$ of only two variables, one function of one variable (from the integration of $h$), and one constant $c$. The symmetry algebra being relatively small in this case, the extension does not become unique anymore (and despite the fact that the given non-vanishing $H$ would eliminate ambiguities of the sort of the previous example). For example, in addition to the action (\ref{eq:HPSM}) for the above data, we can add terms of the form $\int_\Sigma f(X^1) A_2 \wedge A_3$ for an arbitrary function $f$, which corresponds to a modified bivector $\Pi \to \Pi + f(X^1) 
 \partial_2 \wedge \partial_3$. Remarkably, this change of the bivector can be made undone by means of a diffeomorphism on $M$ (leaving $H$ invariant) \emph{provided} $f$ is vanishing at $x^1=0$ (by an $x^1$-dependent shift of $x^2$, as one may easily check); this is, however, no more the case for $f(0)\neq0$. Thus, there exist examples, where the above-mentioned extension has inequivalent solutions. It turns out that this type of ambiguity can be eliminated with an appropriate lift and extension of the symmetry algebra.

Having observed that the symmetry algebra $\g$ can be too small in some cases so as to fix the equivariant extension, we search for an enlargement of it, while still staying inside the context of E-equivariant cohomology with the conditions (\ref{eq:Eequiv}). 
Dropping the condition (\ref{eq:theta}) for a moment, the symmetry algebra consists of sections of the Dirac structure $D \subset TM \oplus T^*M$, which, in particular, is a Lie algebroid. 

Given any Lie algebroid $E$, its Lie algebra on $\Gamma(E)$ can be recovered from the Q-derived bracket (\ref{eq:der}) by means of the degree -1 vector fields $\epsilon$ on $\CM=E[1]$ (which are easily seen to correspond to the sections of $E$). This Lie algebra can be lifted faithfully to $\tCM=T[1]\CM$ by means of the Lie derivative, using again the derived bracket construction (\ref{eq:der}), but replacing $\epsilon$, $\epsilon'$ by $\CL_\epsilon$,  $\CL_{\epsilon'}$, and $Q$ by $\tQ=\rd +\CL_Q$. We may now search an extension of the form $\te=\CL_\epsilon+ \ldots$ for (\ref{eq:Eequiv}). For this we have the following\\

\vspace*{-1em}
\noindent\textbf{Proposition.} Let 
$E\equiv (E, \rho \colon E \to TM, [ \cdot , \cdot]_E)$ 
be a Lie algebroid. The embedding $\CL_\cdot$ of its Lie algebra $(\Gamma(E), [ \cdot , \cdot]_E)$ into the degree -1 vector fields on $T[1]E[1]$ (equipped with the derived bracket) has a unique maximal extension ${\cal{G}}_E$, which is a semi-direct product:
\begin{equation}
0 \to (\Gamma(E \otimes T^*M), [ \cdot , \cdot ]_\rho \to ({\cal{G}}_E,[\cdot, \cdot ]) \stackrel{\stackrel{\CL_\cdot}{\leftarrow}}{\to} (\Gamma(E),[\cdot , \cdot]_E) \to 0\, ,
\end{equation}
where the Lie algebra on $\Gamma(E \otimes T^*M)$ is induced by a pointwisely defined Lie algebra on $E \otimes T^*M \cong \mathrm{Hom}(TM,E)\ni \gamma, \gamma'$, $[\gamma, \gamma']_\rho = - \gamma \circ \rho \circ \gamma'+ \gamma' \circ \rho \circ \gamma$, and the action of $\epsilon \in \Gamma(E)$ on $\gamma = \gamma^a \otimes e_a \in \Gamma(T^*M)\otimes \Gamma(E) \cong  \Gamma(E \otimes T^*M)$ is given by $\epsilon \cdot \gamma := \left(\CL_{\rho(\epsilon)} \gamma^a\right) \otimes e_a + \gamma^a \otimes [\epsilon,e_a]_E$.\\

\vspace*{-0.3em}
\noindent \textbf{Proof} (Sketch)\textbf{.} 
While the derived bracket satisfies a Jacobi-type of equation (its regular left representation satisfies a Leibniz rule with respect to the bracket), it is not automatically antisymmetric (cf., e.g., \cite{Kosmann-Schwarzbach}); for this one needs that $[\tQ,[\te,\te']]$ vanishes for all $\te, \te'\in\CG_E$. Since this has to hold for all $\epsilon,\epsilon' \in \Gamma(E)$ (with the usual identification), this implies that the permitted $\te$ are of the form: $\te = \CL_\epsilon + \theta^i \gamma_i^a(x) \frac{\partial}{\partial \psi^a}$, where $\theta^i= \rd x^i$ and $\psi^a=\rd \xi^a$ are the new ``tangent'' coordinates on $T[1](E[1])$ of degree 1 and 2, respectively. (Note that on a supermanifold $E[1]$ for a Lie algebroid $E$ there are no degree -2 vector fields so $[\epsilon,\epsilon']$ vanishes there for degree reasons; this is no more the case on $T[1]E[1]$, where the degree -2 vector fields are spanned by $ \frac{\partial}{\partial \psi^a}$.) The remainder is a straightforward calculation using the derived bracket $[[\te,\tQ],\te']$ and identifying $\gamma_i^a$ with the components of a section $\gamma$ in $T^*M \otimes E$ for a basis $\rd x^i \otimes e_a$.  $\square$

We now specialize this algebra to a Dirac structure, $E=D$, using also its particularities as compared to general Lie algebroids. For the gauging of a closed 3-form $H$, the algebra $\CG_D$ is still too big. First, we certainly need to impose the constraint (\ref{eq:theta}) on $\epsilon$ of rigid invariance of $H$. But now, in addition to $\epsilon \in \Gamma(D)$, also $\gamma\in \Gamma(D \otimes T^*M)$ acts on $\tH$ 
and we need to restrict it as well. In the present context we always considered Dirac structures in \emph{split} exact Courant algebroids (cf, e.g., \cite{Kotov-Schaller-Strobl} for an explanation of this terminology), i.e.~in addition to the projection $\rho \colon D \to TM$, we also have a projection $\tau \colon D \to T^*M$ (horizontal projection in Fig.~1). For any Dirac structure $D$ we can thus define in a canonical way
\begin{equation} \label{eq:tildeg}
\widetilde{\gamma} := (\tau \otimes \mathrm{id}) \circ \gamma \in \mathrm{Hom}(TM,T^*M) \cong \Gamma(T^*M \otimes T^*M) \, .
\end{equation}
Denote by $\widetilde{\gamma}^A \equiv \frac{1}{2} \widetilde{\gamma}_{ij} \rd x^i \wedge \rd x^j$ its antisymmetric or 2-form part. 
A restriction on $\gamma$ that defines a Lie subalgebra and that turns out to fulfill its purpose is given by the following:
\begin{equation}\widetilde{\g} :=  \{ (v,\a,\gamma)\in \CG_D | \iota_v H = \rd \a \, , \widetilde{\gamma}^A =0\} \, . \label{eq:tildeg2}
\end{equation} 

Before pronouncing the main result of the present paper, we specialize the above Lie algebra to the case of $D=D_\Pi$ of a twisted Poisson structure $(\Pi,H)$.
In that case the map $\tau \colon D \to T^*M$ is an isomorphism, which we have used already to identify $D$ with $T^*M$ in fact, and the anchor map $\rho$  corresponds to simply $\Pi^\sharp \colon T^*M \to TM$ (cf.~also Fig.~1). The projection $\tau$ being an isomorphism permits to identify $\gamma$ with $\widetilde{\gamma}$ in this case, $\gamma \in \Gamma(T^*M \otimes T^*M)$. Then the restriction on $\gamma$ in (\ref{eq:tildeg}) just requires it to be a symmetric tensor, $\gamma \in \Gamma(T^*M \circledS T^*M)=:\Gamma(S^2 \,T^*M)$. Any antisymmetric matrix $\Pi$ induces a Lie bracket on the symmetric matrices $A$, $B$, by means of $[A,B]_\Pi := -(A \, \Pi\, B - B\, \Pi \,A)$. It is this Lie algebra that extends the previously found Lie algebra $\g$ of (\ref{eq:g}) in the form of a semidirect product, with the action of $\g$ on the symmetric tensors just being given by the Lie derivative with respect to  $v=\Pi^
 {\sharp}(\alpha)$, i.e.~$\alpha\cdot \gamma = \CL_{\Pi^\sharp(\alpha)} \gamma$. The semi-direct product thus corresponds to the following exact sequence in this case 
\begin{equation}\label{eq:S2T*}
0 \to (\Gamma(S^2 \,T^*M),[ \cdot , \cdot ]_\Pi)\to
 \widetilde{\g} \to  (\{ \a\in \Omega^1(M) | \iota_{\Pi^\sharp(\alpha)} H = \rd \a \}, [ \cdot , \cdot ]_{D_\Pi})\to 0 \, ,
\end{equation}
where $[ \alpha , \beta ]_{D_\Pi} = 
\rd \iota_{\Pi^\#(\alpha)}\beta + \iota_{\Pi^\#(\beta)}\iota_{\Pi^\#(\alpha)}H$  is the Lie algebroid bracket induced on $T^*M$ by the twisted Poisson structure when restricted to set of 1-forms satisfying (\ref{eq:theta}). As a vector space thus $\widetilde{\g} = \Gamma(T^*M \oplus T^*M \circledS T^*M)\cong \Omega^1(M) \oplus \Gamma(S^2(T^*M))$, the 1-forms are equipped with the above Lie algebroid bracket, the symmetric tensor with the pointwisely defined Lie algebra induced by $\Pi$, and the 1-forms act on them by means of the Lie derivative w.r.t.~their anchor projection $\Pi^\sharp$.

With this choice (\ref{eq:tildeg}), represented by particular degree -1 vector fields on $T[1]D[1]$ as explained above, one finally obtains the following\\

\vspace*{-0.5em}
\noindent \textbf{Theorem.}
Let $H$ be a closed 3-form on $M$ and $D$ a Dirac structure on $(TM \oplus T^*M)_H$ such that the pullback of $H$ to a dense set of orbits 
of $D$ is non-zero. Then the $\widetilde{\g}$-equivariantly closed extension $\tH$ (\ref{eq:Eequiv})  
of $H$ is unique and $\int_{\widetilde{\Sigma}} \widetilde{f}^* (\tH)$ yields 
the (metric-independent part of) the Dirac sigma model \cite{Kotov-Schaller-Strobl} on $\Sigma = \partial {\widetilde{\Sigma}}$.\\

\vspace*{-.3em}

\noindent \textbf{Proof.} 
The proof is a rather straightforward, somewhat lengthy calculation. Let us present its major stages.
For a convenient description of couples $(v,\alpha)$ forming a Dirac structure one may use the identification of $D$ with $TM$ induced by a metric $g$: Such a metric permits the identification $TM \oplus T^*M \cong TM \oplus TM$ and the introduction of the  eigenvalue subbundles 
$E_\pm = {\{ v \oplus \pm v \}}$
of the involution $(v,\alpha) \mapsto (\alpha,v)$. Clearly, $E_+$ as well as $E_-$ can be 
identified with $TM$ (projection to the first factor). 
\begin{figure}[b] \centering
\includegraphics[width=0.8\linewidth]{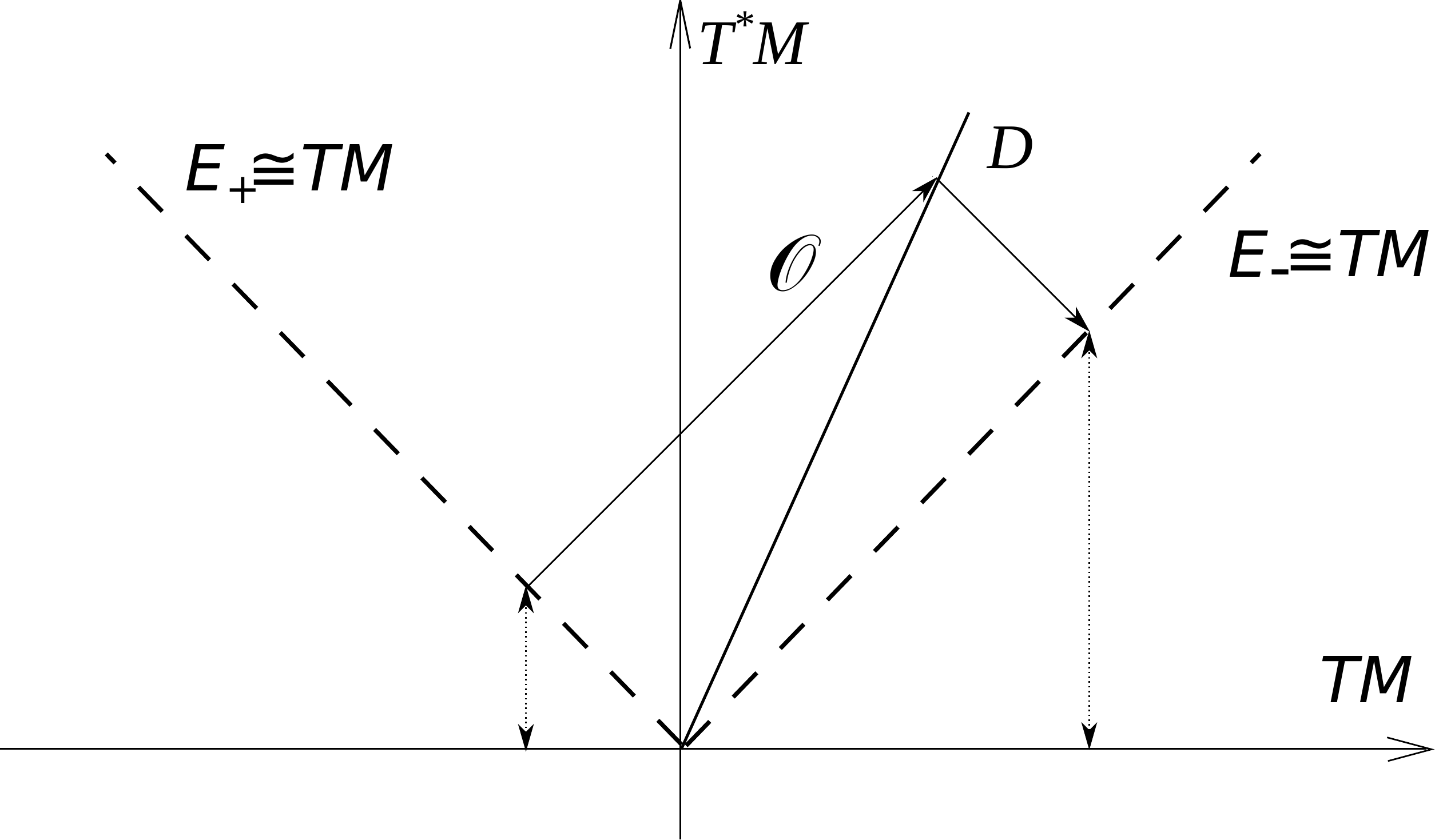}
\caption{\label{fig:graphO}  Dirac structure as a graph of 
an orthogonal operator ${\cal O} \in\Gamma(\mathrm{End}(TM))$. While with respect to the frame $T^*M/TM$ only (twisted) Poisson and presymplectic structures can be written as a graph, cf.~figure \ref{fig:Dirac}, the frame $E_+/E_-$, introduced by means of an auxiliary metric, permits to write \emph{every} Dirac structure $D$ as a graph (since necessarily $D\cap E_{\pm} = 0$).}
\end{figure}
In this setting any Dirac structure corresponds to 
a point-wise orthogonal operator ${\cal O} \in\Gamma(\mathrm{End}(TM))$, viewed as a map from $E_+\cong TM$ to $E_-\cong TM$
(see figure \ref{fig:graphO}, as well as \cite{Kotov-Schaller-Strobl} for the full argument), 
subject to the following (twisted Jacobi-type) integrability condition
\begin{equation}\label{eq:gJac}
g\left({\cal O}^{-1}\nabla_{(\mathrm{id}-{\cal O})\xi_1}({\cal O})\xi_2, \xi_3\right) + cycl(1,2,3) = 
  \frac{1}{2} H ((\mathrm{id}-{\cal O})\xi_1, (\mathrm{id}-{\cal O})\xi_2, (\mathrm{id}-{\cal O})\xi_3 ).
\end{equation}
If the operator $(\mathrm{id} +{\cal O})$ is invertible, this corresponds precisely to the Dirac structure $D_\Pi$ of a twisted Poisson structure 
with bivector $\Pi = (\mathrm{id} - {\cal O})\cdot (\mathrm{id} + {\cal O})^{-1}$ (cf.~also \cite{Courant} and figure \ref{fig:Dirac}); in this case, the condition (\ref{eq:gJac}) can be rewritten as  $\frac{1}{2} [\Pi, \Pi] = \langle H,\Pi^{\otimes 3} \rangle$. Such as for the graph of a bivector $\Pi$ we could parametrize the elements of the corresponding Dirac structure $D_\Pi$ by elements $\alpha \in T^*M$ (cf.~figure \ref{fig:Dirac}), now for an arbitrary Dirac structure $D$ we can parametrize its elements by vectors $w \in TM$; then the pair $(v,\alpha) \in TM\oplus T^*M$ takes the form $(w - {\cal O}w, g(w + {\cal O}w,\cdot))$.

For the equivariant cohomology construction we use the above identification of $D$ with a tangent bundle, equipped with a non-standard Lie algebroid structure: so, $D[1] \cong T[1]M$ -- where again we declared fiber-linear coordinates to be odd of degree $+1$.
The $Q$-structure is of the form (\ref{eq:Q}) with the 
anchor map $\rho = (\mathrm{id} - {\cal O})$ 
(cf.~the end of the previous paragraph) and the structure functions $C^a_{bc}(x)$ being induced by the Courant-Dorfmann bracket (\ref{eq:Dorfmann}), which 
explicitly are given by eq.~(40) in \cite{Kotov-Schaller-Strobl}. 

As explained above the target $Q$-manifold of the sigma model 
is the shifted tangent bundle to $D$, $\widetilde {\cal M} = T[1]D[1]$, equipped 
with the $Q$-structure $\tilde Q = d +\CL_Q$, and the algebra of symmetries 
encoded in the degree $-1$ vector fields $\tilde \varepsilon$ is given by $\tilde \g$ defined by eq. (\ref{eq:tildeg}), cf. the Proposition above. 

Now starts the purely computational part of the proof, performed in some local coordinates. One starts with the most general superfunction $\tilde H$ on $T[1]D[1]$ of total degree $3$, the degree matching the dimension of $\tilde \Sigma$, and applies the conditions (\ref{eq:Eequiv}) to this ansatz, where the first of them can be taken account of directly in the ansatz. Note then that $T[1]D[1]$ is naturally equipped with a double $\mathbb Z$-grading: the first one coming from the shift in $D[1]$, called ghost-number previously, and the second one from the tangent bundle and being the form degree (of the differential form living on $D[1]$). Thus, a superfunction of total degree $3$ contains six types of terms: 
compare for example $\eta \rd x^1 \rd x^2$ and $\rd \eta \rd x^1$, where $x$'s are coordinates on $M$ and $\eta$ a fiber-linear coordinate on $D[1]$. 
Since the differential $\tilde Q$ is of total degree $1$ and mixes the double grading, 
the condition $\tilde Q \tilde H = 0$ determines one half of the (degree $(0,0)$) 
coefficients in terms of the other half. 

It then remains to solve the conditions $\tilde \varepsilon \tilde H = 0$. 
The vector field $\tilde \varepsilon$ is lowering the total degree 
by $1$. Consequently, $\tilde \varepsilon \tilde H$ contains four types of terms 
which have to vanish independently. It is useful here to first consider only the extension part of $\tilde \g$, i.e.~the vector fields parametrized by $\tilde \gamma$
with $\varepsilon \equiv 0$:  
the terms of form-degree $1$ not containing coordinates of total degree $2$ (of the form $f(x)\eta \rd x$, but not proportional to $d\eta$)
enforce a relation between two groups of coefficients out of three, left 
when restricting to $\tilde H$ being $\tilde Q$-closed; the relation 
on the terms of form-degree $2$ determines these coefficients up to 
a global prefactor. One then is left with establishing the relation of those terms 
to the form-degree $3$ term in $\tilde H$, which is nothing but $H$ due to the 
first condition in (\ref{eq:Eequiv}). This is done by considering 
the unextended part of $\tilde \g$, i.e.~vector fields $\widetilde \epsilon = {\cal L}_\epsilon$ parametrized by 
$\g$: Due to the condition that we impose on the pullback of $H$ to
the orbits of $D$, ensuring that $\iota_v H \neq 0$, the equation (\ref{eq:theta}), which enters due to the definitions (\ref{eq:tildeg2}) or (\ref{eq:g}), is satisfied for some non-vanishing $\rd \alpha$. This, together 
with the relation on the form-degree $2$ terms of $\tilde \varepsilon \tilde H$, determines the above-mentioned prefactor and implies the uniqueness of the extension.
One can verify (using the orthogonality of ${\cal O}$ and the integrability 
condition (\ref{eq:gJac}) on it) that the result is compatible with the conditions coming from the other form-degrees, furthermore.
 
The remaining part is a straightforward computation of $\widetilde f^*(\tilde H)$
as described in the second half of section \ref{sec:BRST}. Performing partial integration of the result using Stokes' theorem, 
one obtains the (topological part of the) functional of the DSM in the following form: 
\begin{equation} \label{eq:DSM}
   S^0_{DSM} = \int_{\Sigma} g( \rd X\stackrel{\wedge}{,} (\mathrm{id}+{\cal O}) {\cal A}) + g({\cal A}\stackrel{\wedge}{,}{\cal O} {\cal A}) + \int_{\widetilde \Sigma} H \, . 
\end{equation}
The full action functional of the DSM of \cite{Kotov-Schaller-Strobl} results from adding a kinetic term with minimal coupling, $S^{\lambda}_{kin} = \frac{\lambda}{2} 
\int_\S g((\rd X - V) \stackrel{\wedge}{,} *(\rd X - V))$ with $V\equiv {\cal A}-{\cal O} {\cal A}$; $S_{DSM}\equiv S_{DSM}^{\lambda}= S^0_{DSM} +S^{\lambda}_{kin}$.
 But, as mentioned in the introduction, this term only serves as a regulator, not influencing the ``physics'' of the model otherwise, which results from the limit $\lambda \to 0$ corresponding to (\ref{eq:DSM}) already.

\begin{figure}[htp] \centering
\includegraphics[width=0.5\linewidth]{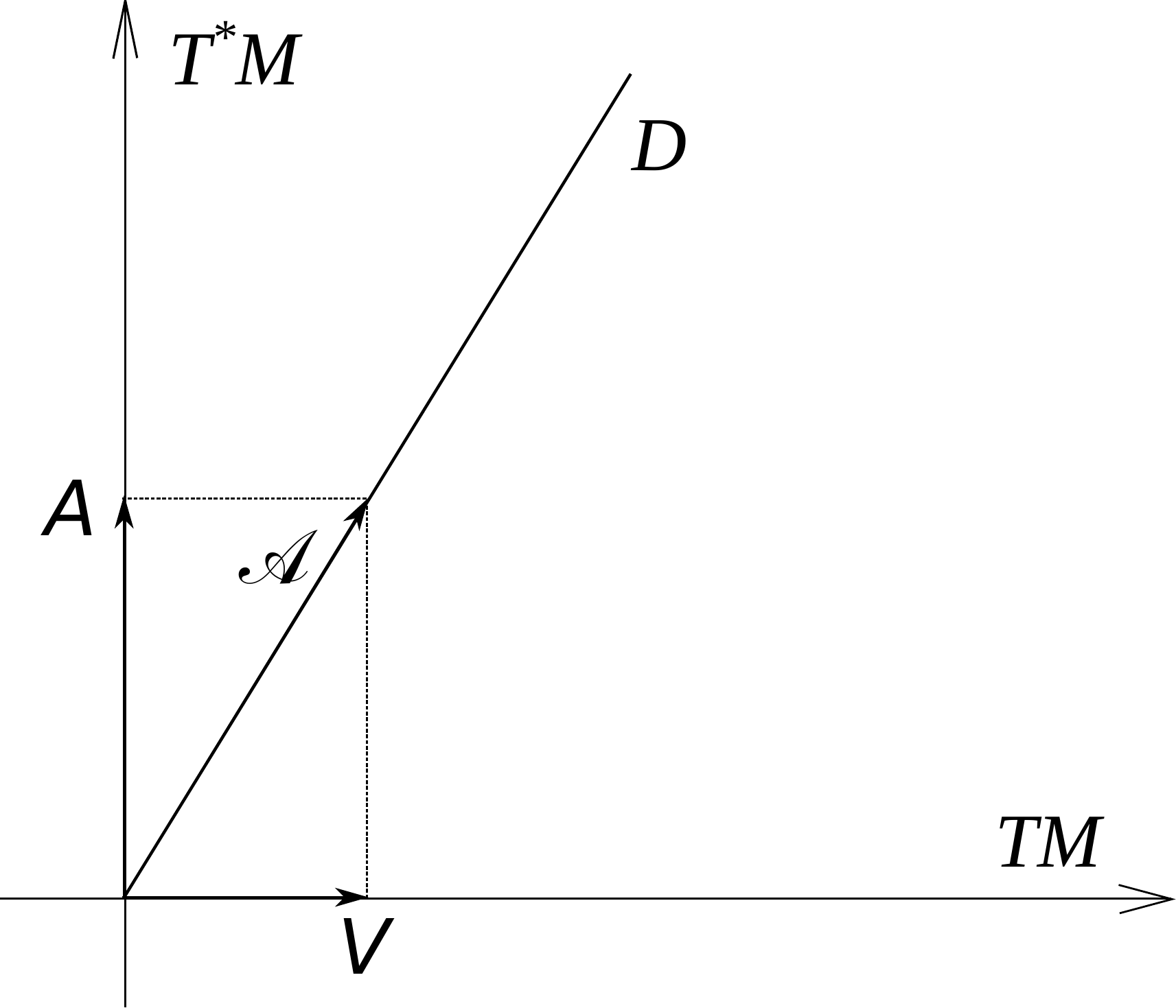}
\caption{\label{fig:DS}  Schematic picture of the field content of the Dirac sigma model. Beside the map $X \colon \Sigma \to M$, there are 1-form fields taking values in the Dirac structure $D$ chosen to define the model: ${\cal A} \in \Omega^1(\Sigma, X^* D)$. The projection of ${\cal A}$ into the $TM$ and $T^*M$ directions are called $V$ and $A$, respectively. They are not independent fields, however. The complete field content of the Dirac sigma model combines into vector bundle maps from $T\Sigma$ to the subbundle $D \subset TM \oplus T^*M$ defined by the Dirac structure.} 
\end{figure}

Despite its appearance, the functional (\ref{eq:DSM}) does not depend on the auxiliary metric at all. This becomes evident by rewriting it into the form:
\begin{equation} \label{eq:DSMtop}
S^0_{DSM} = \int_{\Sigma} A_i \wedge \rd X^i - \frac{1}{2} A_i \wedge V^i + \int_{\tilde{\S}}  \tilde{X}^* H \,,
\end{equation}
with $A_i \equiv g_{ij}(X) ({\cal A}-{\cal O} {\cal A})^j$ and $V^i\equiv {\cal A}^i-{\cal O}(X)^i{}_j {\cal A}^j$. The price to pay here is that the variables $A$ and $V$ are not independent, but are just the $T^*M$- and $TM$-components of the independent gauge field along the Dirac structure, as illustrated in figure \ref{fig:DS}. This becomes particularly evident in the case of the Dirac structure $D_\Pi$, where $V^i = \Pi^{ji}(X)A_j$ (cf.~figure \ref{fig:Dirac}), with $A_i$ being independent 1-form fields, in which case the action (\ref{eq:DSMtop}) evidently reduces to the one of the twisted Poisson sigma model (\ref{eq:HPSM}).
$\square$\footnote{For some more explicit formulas and a detailed presentation of the above procedure for the twisted Poisson sigma model the reader may consult also \cite{Vladimir-PhD}.}

We briefly remark that such as any Lie algebroid $E \to M$ also a Dirac structure induces a (possibly) singular foliation on $M$ into orbits (integration of the image of the anchor map $\rho \colon E \to TM$). For the case of a Dirac structure $D$ these turn out to be $H$-twisted symplectic. The condition that the pullback of $H$ to any (non-trivial) leaf is non-zero excludes ambiguities like those encountered in the Poisson sigma model (\ref{eq:action}). We stress that it is sufficient that the pullback is non-zero \emph{somewhere} on such a leaf, not non-zero everywhere on it. The extension from $\g$ to $\widetilde{\g}$, on the other hand, excludes ambiguities of the sort mentioned in the example following the one of the PSM. 

The Dirac sigma model contains a metric independent part which for a twisted Poisson structure $(\Pi,H)$ is precisely given by (\ref{eq:HPSM}). The metric dependent part only serves as a kind of regulator and it consists simply of a standard kinetic term (like in (\ref{eq:sigma})) minimally coupled. 

\newpage
\section{Summary and Outlook}
The main result of the paper is that a BRST-type adaptation of standard equivariant cohomology given by the formulas (\ref{eq:Eequiv}), applied appropriately to twisted Poisson and, more generally, Dirac structures, can explain the form of the twisted Poisson sigma model and the (essential part of the) Dirac sigma model, respectively. The main idea for this formulation of equivariant cohomology goes back to \cite{Kotov-Strobl}, but we adapted it by the lift from $\g$ to $\widetilde{\g}$, which was essential for the uniqueness result. 

We have chosen conditions for the main result, the above Theorem, that are sufficient for uniqueness. It may well be possible to relax them keeping a unique extension. In fact, we know examples like of an $H$-twisted Poisson structure $\Pi$ on a three-dimensional manifold $M$, which necessarily violates the condition since the leaves are even dimensional, where one still gets uniqueness of the extension problem. Also it may be interesting to parametrize the ambiguities in the general case (or under weaker assumptions). 

In this article we focused mainly on a careful formulation of the extension problem, making as clear as possible its ingredients and prerequisites. It is in principle also interesting to explain in some detail, how this indeed yields a gauge invariant functional and how the infinitesimal gauge transformations can be described as inner automorphisms of an appropriate bundle. For this we may, however, refer to \cite{Bojowald-Kotov-Strobl,Kotov-Strobl} and a short paper \cite{Salnikov-Strobl2} parallel to this one where we focus on the novelties that algebroid methods in general offer even in conventional contexts such as the one of standard gauging problems. We remark as an aside in this context that the 
extension part of (\ref{eq:S2T*}) by symmetric tensors $\gamma$ permit to recover  precisely trivial gauge symmetries  (in the sense of, e.g., \cite{Henneaux-Teitelboim}). Moreover, one may replace the two separate conditions on $(v,\alpha)\in \Gamma(D)$ and $\gamma$ given by (\ref{eq:theta}) and (\ref{eq:tildeg}), respectively, by a single one combining the two while still keeping the topological part of the Dirac sigma model gauge invariant. This can have some technical advantages in some contexts. Also the presence of the kinetic term deforms/changes the gauge symmetries. We may come back to these issues in more detail elsewhere. 

Finally, we remark that there were several attempts in the mathematics literature to define Lie algebroid equivariant cohomology, cf, e.g., \cite{Roubtsov, algd_coh, algd_coh2}. 
We believe that for general Lie n-algebroids there should be a formulation that contains the eqs.~(\ref{eq:Eequiv}). On the other hand, the details of the additional structures (given by the Dirac structure and the splitting) needed to obtain the results of the present paper may add another class of examples to the general development of a theory of algebroid equivariant cohomology that one may want to reproduce in a more general context.

\vspace{1cm}
\acknowledgments
We gratefully acknowledge inspiring discussions with A.~Alekseev, A.~Kotov and V.~Roubtsov.


\end{document}